# IT Governance in Actor-Network Mode of Collaboration : Cost Management Process Based on Game Theory

Mohammed Salim BENQATLA[1], Dikra CHIKHAOUI[2] and Bouchaib BOUNABAT[3]

[1] Al-Qualsadi Research & Development Team National
Higher School for Computer Science and Systems analysis (ENSIAS) Med V – Souissi University, Rabat, Morocco

[2] Al-Qualsadi Research & Development Team National
Higher School for Computer Science and Systems analysis (ENSIAS) Med V – Souissi University, Rabat, Morocco

[3] Al-Qualsadi Research & Development Team National
Higher School for Computer Science and Systems analysis (ENSIAS) Med V – Souissi University, Rabat, Morocco

**Abstract**
IT governance, like global governance of projects, requires cooperation between several actors. In general, such cooperation builds a collaboration network between entities. Many works in the literature interested in collaboration network, but no one of them were focused directly on how to build a network in an IT governance perspective. In this paper, we investigate how game theory can be exploited to provide a formal implementation of Cost Management Process, while highlighting Actor-Network as a framework of collaboration and its incentive stage as a key step for network construction. Our objective is to propose an approach of network establishment, by inciting actors through cost savings. For that, we use Shapley Value to answer the question: For the sake of IT governance, which coalitions are likely to form in order to ensure best cost-saving objectives in ANT mode of collaboration? A graphical tool is developed also to visualize and simulate networks evolution.

## 1. Introduction

IT governance is a business function that relies on people to manage technological resources. Therefore, it is susceptible to the games they play and their consequences [6], as it happens during bargaining and cost sharing negotiations. In order to take advantage of their benefits, companies must promote the adoption of concepts and methods that favor the effective use of technological resources [5]; they are looking to best practice frameworks such as COBIT to improve the quality of their IT governance.

On another side, collaborative network has become a key enabler for trading success and economic growth. Within IT governance, cooperation takes different forms, from simple information exchange, to business processes interoperability among independent enterprises [6] [11], and also in term of cost-sharing. In fact, independent businesses become able to collaborate in order to have benefic results for all [2].

In this paper we investigate how IT governance can benefit from the use of game theory to implement Costs Management process. The Literature review shows that so far no publication has addressed this subject directly. Therefore, connecting game theory to IT governance seems to be an open road for research and publication [1].
In our context we use Shapley value [10] as a fair cost sharing solution to divide costs between actors involved in IT project, relative to their marginal contributions.
We introduce Actor-Network Theory (ANT) [3], [4] as a framework of collaboration which helps us make sense to interaction evolution between different actors of business network. The paper is organized as follows: Section 2 presents a review of IT governance initiatives with Game Theory. Section 3 presents the different concepts and theories used. Section 4 introduces the proposed cooperative network building game. Section 5 presents numerical application that calculates shared-cost between actors using Shapley Value in a realistic Actor-network context via a developed java platform. Finally, Section 6 concludes this paper.

## 2. IT Governance Vs Game Theory - Literature Review

We provide in this section a brief literature review of game theory as a support in IT governance initiatives. In a Systematic literature Review [1] six papers was selected to aid the implementation of relevant aspects of governance using game theory. For example, the use of game theory to support the governance of common resources is approached in [27]. The prevention of environmental accidents and increasing readiness for action in the aftermath of accidents with the support of game theory and governance is discussed in [7]. The use of game theory to improve the planning of organizational change is dealt with in [8], and the use of game theory to improve the efficacy of governance committees in knowledge alliance is studied in [9]. The selected papers indicate that game theory could be used in similar ways to aid IT governance initiatives, but none of them deal with the governance of information technology directly. Therefore, the use of game theory to support IT governance is actually an open road for research and publication [1]. In fact, while involved in IT governance activities, actors are likely to play cooperative games.

## 3. CONCEPTS

### 3.1 COBIT

ISACA [14] develops and maintains the internationally recognized COBIT framework, helping IT professionals and enterprise leaders fulfill their IT Governance responsibilities while delivering value to the business.
COBIT (Control Objectives for Information and Related Technologies), now in its fifth edition released in April 2012, describes a set of good practices for the board, executive management, and operational business and IT managers. It helps organizations to create value from IT by maximizing the benefits and minimizing the risks associated with IT, ensuring that IT meets the legislative and regulatory requirements, and achieves alignment of IT strategy with business goals [12].
COBIT integrates IT governance into enterprise governance and covers the functions, processes and services across the enterprise, both internal and external.
The COBIT 5 processes are split into governance and management "areas". These 2 areas contain a total of 5 domains and 37 processes, Governance of Enterprise IT[Evaluate, Direct and Monitor (EDM)] and Management of Enterprise IT [Align, Plan and Organise (APO) ; Build Acquire and Implement (BAI) ; Deliver Service and Support (DSS); Monitor Evaluate and Assess (MEA) ] [14].

We focus our work on the sixth process of APO domain; Budget and Cost management process.

### 3.2 Actor Network Theory (ANT)

The theory of translation or sociology of translation known as Actor Network Theory: (ANT) was developed as part of research on the innovation process and is rooted in a socio-technical approach to organizations. The founders of this current, Akrich, Callon and Latour [23] have shown that successful innovation depends on the success of unprecedented association between multiple and different actors. From this association, mobilization and cooperation of all stakeholders will emerge a socio-technical network and a dynamic production that aim process efficiency and success.
The second important notion of ANT is the "Actant" Callon and Latour borrow this concept to semiotician Greimas. The latter replaces the term personage by the term actant, that "who does or endure an act", because it applies not only to humans but also to animals, objects, concepts. The actants may be human or non-human and should be treated with the same importance as required by the principle of symmetry.
In order to reach a step of construction of a network, Callon and Latour defined an approach, inspired by ethnomethodology [27], which bears on a sequence of steps called the translation sequence. To translate is to "express in his own language what others say and want, to set up as spokesman" [3], but translate it is also, negotiate, perform a series of movements of all kinds and this to each sequence of the process, which can be defined in four main steps:

1. Problematization:

"The problematization or how to become essential?", "The problematization, as its name indicates asking at first a problem. This is to raise awareness to a number of actors that are concerned by this problem, and that everyone can find satisfaction through a solution that translators are able to offer" [17], so problematization is the effort made by the actors to convince that they have the right solution[16]. It "describes a system of alliances or associations between entities, defining this, [their] identity and what they want" [18].

2. Interessement

"The incentive devices or how to seal alliances", the incentive is in fact for Callon "all actions through which an entity is trying to impose and stabilize the identity of the other players who is defined in problematization" [17] incentive is the second phase, consists of "deployment speeches, objects and devices intended to attract and attach different players to the Network" [19].
It is building the interface between the interests of different actors and the strengthening of the relationship between these interests. In the area of strategy, it can be a system of alliances to ensure that the different members of the organization are involved in the strategic process.

The main thing is to translate the interests of other actors in order to get them to take part in the network. To translate the interests of others, we can either convince them that there are common interests and that the proposed solution also serves their interests or manipulate their interests and objectives or finally become unavoidable.

3. Enrollment

"How to define and coordinate the roles", Enrollment is "the set of multilateral negotiations, beatings forces or tricks that come with sharing and allow it to succeed" [17]. For enrollment, each actor in the network is assigned a role. This role is related to the translation of their interests. For Callon, «the enrollment is to describe the set of multilateral negotiations, coups or intelligence accompanying sharing and allow it to succeed" [18] .The enrollment can thus be regarded as stabilizing the system of alliances set during the phase of the incentive. This system is the result of multilateral negotiations, trials of strength and stratagems [18]. It is during this phase to confront showdowns integrating new actors to the networks or by strengthening links between network members.

The enrollment phase is the key to the success or failure of innovation [18], but this phase is not studied formally in the literature on control.

4. Mobilization

Last phase of translation, the mobilization is to gather its allies. It is the cockpit of the various interests in a way that they remain more or less stable [20], it raises the question of the representation of stakeholders and enrolled in the project which is then established as spokespersons of the groups they represent [21]. However, "everyone can act very differently to the solution proposed: the abandon, accept it as it is, change the modalities which accompany or statement that it contains, or even they will be appropriated in the transferring in a completely different context" [18].

In a particular way, incentive phase of ANT can be analyzed from a cooperative game with transferable utility point of view. Our objective is to set up the network by incenting actors through cost savings. For that, we use Shapley Value to answer the question: Which coalitions are likely to form in order to ensure best translation of cost-saving objectives in an actor-network context?

3.3 Cooperative games theory

The cooperative game theory can be applied to the case where actors can achieve more benefit by cooperating than staying alone, it consists of two elements: (i) a set of players, and (ii) a characteristic function specifying the value created by different subsets of the players in the game [24]. The coalition formation problem is one of the important issues of game theory, both in cooperative and non-cooperative games. There are several attempts to analyze this problem. Many papers tried to find stable coalition structures in a cooperative game theoretic fashion. If we suppose that forming the grand coalition generates the largest total surplus, it is natural to assume that the grand coalition structure will eventually form after some negotiations [26]. Then, the worth of the grand coalition has to be allocated to the individual players, according to the contribution of each player [26]. We are interested in this work in cost-sharing between coalition members likely to form using Game theory as a device for ANT interessement stage.

3.4 Cost management process in a business collaboration network

A major concern of senior management is the level of the IT costs and their recovery. Implement a cost management process consists on comparing costs to budgets. Stakeholders are consulted to identify and control the total costs and benefits within the context of the IT strategic and tactical plans, and initiate corrective action where needed [30]. The process promotes partnership between the different actors; enables the rational use of IT resources; and provides transparency and accountability.

Collaboration network has created a need for new management control practices. Collaborative cost management process is defined as buyers' and suppliers; coordinated efforts to reduce costs [15]. The theoretical literature on interorganizational relationship formation is fragmented, with several disciplines contributing to the field. Transaction cost economics (TCE) [26], actor network theory [27], industrial network approach [28], and structuring theory [29] are the dominant theoretical perspective of interorganizational setting as well as interorganizational cost management research.

In this paper we focus on actor network theory approach and deal with cost management both as a process of IT governance and as a mechanism of interessement.

## 4. Actor-Network building Game

In our framework players are actors of network. To the extent that they may have common interests, actors are required to cooperate in advance to take and implement joint decisions, coordinate their actions and pool their winnings & cost. It appears a cooperative game where the actors come together to form coalitions, and all of whom seek to optimize the quality and cost of their own operations. They can, through cooperation, realize gains in the form of cost reduction. We can discuss it during the game in terms of the distribution of costs rather than gains.

This is the approach taken here. Then costs are divided between the players relative to their marginal contributions. To formalize the cost-sharing model with cooperative game in this coalition building process, we apply a concept of axiomatic solution, in this case the Shapley value.

Let $N = \{1, \ldots, n\}$ be a finite set of players. A coalition is any subset of N. The set of all coalitions is denoted by $2^n$. A coalitional form concern on a finite set of players $S\{1, \ldots, n\}$ is a function v from the set of all coalitions $2^n$ to the set of real numbers R with $v(\emptyset) = 0$. $v(S)$ represents the total worth the coalition S can get in the game v.

### 4.1 The use of Shapley value

The Shapley value is a very common cost-sharing procedure in cooperative game theory essentially based on the so-called incremental costs [24]. The Shapley value of player i in the game given by the characteristic function V is the share of the surplus should be assign. It's a weighted average of the contributions of player i to reach of the possible coalition.

For example, consider a game with three players, i1, i2 and i3. Assume that player i1 is the first player of the game, i2 is the second player to join the game and player i3 is the last one. Player i1 is allocated a cost C({i1}), player i2 is allocated a cost C({i1, i2}) − C({i1}), and player i3 a cost C({i1, i2, i3}) − C({i1, i2}). The Shapley value assumes that the order of arrival is random and the probability that a player joins first, second, third, etc. a coalition is the same for all players. Assume that forces of each coalition are known in the form of the characteristic function V. The cost allocated to a player i in a game including a set N of players is given by:

$$\phi i(N) = (\sum_{S \subseteq N: i \in S} (\frac{(|S|-1)!(|N|-|S|)!}{|N|!} ([C(S) - C(S \setminus \{i\})]) \quad (1)$$

|N| and |S| respectively, the total number of players and the one belonging to the coalition S.

An alternative equivalent formula for the Shapley value is:

$$\phi i(N) = (\frac{1}{|N|!} \sum_R (v(PRi \cup \{i\}) - v(PRi)) \quad (2)$$

Where the sum ranges over all |N| orders R of the players and $PRi$ is the set of players in N which precede i in the order R.

Choosing a method of cost allocation is not an easy thing. According to the literature Shapley value seems to be suitable to this context of actor-Network building game. In fact, Shapley imposes four axioms to be satisfied (Efficiency, Symmetry, Dummy and Additivity).

(i) **Efficiency**: players precisely distribute among themselves the resources available to the grand coalition. Namely, Efficiency: $\sum i \in N\ \varphi i(v) = v(N)$.

(ii) **Symmetry**: Players i,j ∈ N are said to be symmetric with respect to game v if they make the same marginal contribution to any coalition, i.e., for each S ⊂ N with i,j ∉ S, v(S ∪ i) = v(S ∪ j). In another way if players i and j are symmetric with respect to game v, then φi(v) = φj(v).

(iii) **Dummy**: If i is a dummy player, i.e., v(S ∪ i)- v(S) = 0 for every S ⊂ N, then φi(v) = 0.

(iv) **Additivity**: φ (v+w) = φ (v) +φ (w), where the game v+w is defined by (v+w)(S) = v(S) +w(S) for all S.

The dummy, symmetry (meaning that two players have the same strength Strategic will receive the same gain) and efficiency make the Shapley value particularly attractive for treating the problem of equitable sharing of resources common to several economic agents.

### 4.2 Experimental setup of cost sharing within a public institution with several actors

An administration with several actors/stakeholders (department, partners, suppliers...) may wish to establish a costs management process that encourages collaborators to contribute to minimizing the common cost. As shown Shubik (1962), the allocation of common costs in the company can be seen as a cooperative game between different departments.

To fix ideas, consider the following example with three directions (A, B and C) of the same department that are in agreement with a company to perform backup sites. The project amounts to 10 million for each direction taken separately. For technical reasons, the service provider offers cost (reduced) respectively 16, 17 and 18 for joint contracts between A and B, A and C, B and C. The contract involving the three directions has a cost of 24. The cost function is given then by:

TABLE I. TABLEAU OF COSTS

| Coalition | Cost |
|---|---|
| A | 10 |
| B | 10 |
| C | 10 |
| AB | 16 |
| AC | 17 |
| BC | 18 |
| ABC | 24 |

The construction of a common backup site might be more profitable than building smaller sites. Indeed, the three directions get a fair deal, and are motivated to form a coalition since their cost parts are below their costs of going it alone. How costs should they are distributed among the three directions?

This issue can be described by a three-player game, N = {A, B, C} is thus obtained:

TABLE II. THE CHARACTERISTIC FUNCTION ELEMENTS

| Coalition | Gain |
|---|---|
| A | 0 |
| B | 0 |
| C | 0 |
| AB | 4 |
| AC | 3 |
| BC | 2 |
| ABC | 6 |

Applying Shapley formula (1), there are six possible arrival orders (3!). They are listed in the following table which gives the marginal contributions according to each of them.

For example, $PA(ABC) = v(\{A\}) - v(\theta) = 0-0 = 0$, $PB(ABC) = v(\{AB\}) - v(\{A\}) = 4-0 = 4$, etc.

The distribution of v (N) cost reduction according to the Shapley value is given by $\varphi(v) = (2.5, 2, 1.5)$. In terms of cost sharing, the calculation is illustrated in Table III.

TABLE III. CALCULING SHAPLEY VALUE

| Entry order | Marginal contributions | | |
|---|---|---|---|
| | A | B | C |
| ABC | 0 | 4 | 2 |
| ACB | 0 | 3 | 3 |
| BAC | 4 | 0 | 2 |
| BCA | 4 | 0 | 2 |
| CAB | 3 | 3 | 0 |
| CBA | 4 | 2 | 0 |
| Total | 15 | 12 | 9 |
| Shapley Value | 15/6 | 12/6 | 9/6 |

This means that about 24 million, the directions A, B and C have to pay 7.5; 8 and 8.5 respectively.

## 5. Experimental results

After completing this research, and in order to validate the approach presented in this paper, we developed a java platform composed of two modules; the first one allows to draw network as it is and design the different information about the actor network, the second module permits to calculate actors Shapely value and simulates coalitions costs.

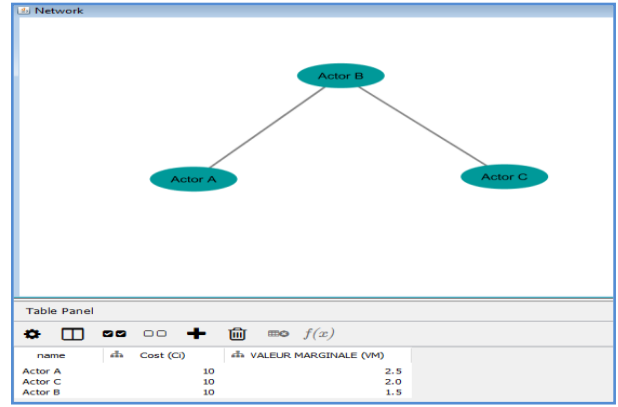

Figure 1. Marginal Values in ABC Coalition

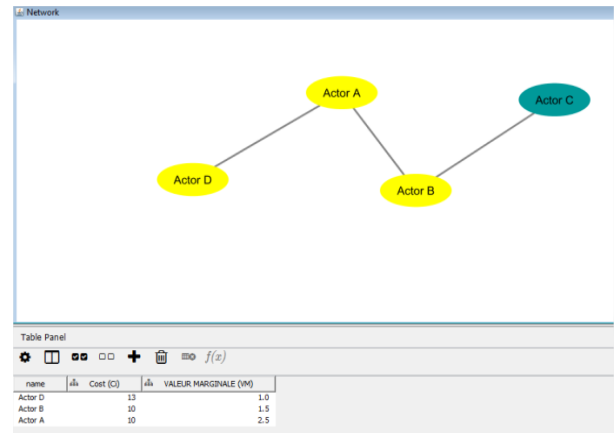

Figure 2. Marginal Values in ABCD Coalition

Numerical results demonstrate that our approach permits to achieve very effective cost allocations, thus representing an efficient framework for the conception of stable networks.

## 6. Conclusion

The build of partnership and coalition intra and inter-departments appears a strategic decision to reduce costs and achieve the submitted projects. This incentive approach could be introduced by the network administrator or the deciders makers in order to increase the users' cooperation level. The rules of sharing common costs and benefits of cooperation are important factors of competitiveness, performance, transparency and motivation, therefore for good governance.

We addressed in this paper It Governance, in particular, the Budget and Cost Management Process from a cooperative game point of view.

The feature of this work is the use of Actor-Network to establish collaborative network, by inciting actors to choose the best coalition through cost saving applying Shapley values. The proposed work is supported by a software tool which enables to design network and calculate actor's Shapley Value.

The main contributions of this work can therefore be summarized as follows:
- Cost sharing as incentive device and formal support of budget and costs management process
- Formulation of the Actor-network building problem as a cooperative game, where players (actors) cooperate to reduce costs
- Implementation of a graphical tool in order to design and simulate the actor-network evolution based on cost calculation approach

Apart from that, our present theoretical model still requires more elaboration on details, and the Shapely value that can be utilized to support interessement stage of ANT remains as a proposal in the case of budget and costs management process. Future work may require more empirical research with different types of actors and objectives.

**MOHAMMED SALIM BENQATLA** is a Software Engineer graduated from ENSA (2007) (National Higher School for Application Science), and is a "Ph.D. candidate" at ENSIAS since 2012. His research focuses on interoperability monitoring within business collaboration networks specially those involving public administration. He is an auditor of system information on the Administration of Customs and Indirect Taxes of the Kingdom of Morocco since 2009. He also oversees the IT operating activity in this department.
**Dikra CHIKHAOUI** is a Software Engineer graduated from ENSA (2007) (National Higher School for Application Science), and is a "Ph.D. candidate" at ENSIAS since 2013. His research focuses on maturity model of governance of information system involving public administration. She is a project manager of system information on the General Treasury of the Kingdom of Morocco since 2009.
**Bouchaib BOUNANABT** PhD in Computer Sciences Professor in ENSIAS, (National Higher School for Computer Science and System analysis), Rabat, Morocco Expert in National ICT Strategies and E-Government to the World Bank, IDB (Islamic Development Bank), ECA (Economic Commission for Africa), UNESCO, ISESCO, UNIDO (United Nations Industrial Development Organization) Candidate of the Kingdom of Morocco for the Secretariat-General of the Arab Information and Communication Technologies Organization, Member of the editorial board of the journal of Information and Communication Technologies and Human Development, Member of the board of Internet Society - Moroccan Chapter